\documentclass[12pt]{article}
\usepackage{epsfig}
\usepackage{amsfonts}
\usepackage{amscd}
\usepackage{latexsym}
\usepackage{amsmath,amssymb}
\usepackage{verbatim}
\usepackage{setspace}
\usepackage{color}
\usepackage{cite}
\usepackage{caption}
\usepackage{url}
\usepackage{graphicx}
\usepackage{tikz}
\usepackage{tikz-3dplot}
\usepackage{cite}
\usetikzlibrary{calc}
\usetikzlibrary{positioning}
\usetikzlibrary{arrows,snakes,backgrounds}
\usetikzlibrary{patterns}
\usetikzlibrary{decorations.pathreplacing,decorations.markings}
\usetikzlibrary{decorations.pathmorphing}
\usetikzlibrary{shapes}

\usepackage[textheight=9in, textwidth=6.5in, letterpaper]{geometry}

\usepackage{hyperref}
\hypersetup{
    colorlinks,
    citecolor=black,
    filecolor=black,
    linkcolor=black,
    urlcolor=black
    linktoc=all
}

\numberwithin{equation}{section}
 
 \def\p{\partial}
 \def\bz{{\bar z}} 
  \def\bw{{\bar w}} 
  \def\ip{${\cal I}^+$}
\def\0{{(0)}}
\def\1{{(1)}}
\def\2{{(2)}}

\def\ci{{\mathcal I}}
\def\co{{\mathcal O}}
\def\cs{{\mathcal S}}

\def\<{\langle }
\def\>{\rangle}
\def\o{{\rm out}}
\def\i{{\rm in}}
\newcommand{\bea}{\begin{eqnarray}}
\newcommand{\eea}{\end{eqnarray}}
\newcommand{\be}{\begin{equation}}
\newcommand{\ee}{\end{equation}}
\newcommand{\ba}{\begin{align}}
\newcommand{\ea}{\end{align}}
\def\be{\begin{equation}}
\def\ee{\end{equation}}
\def\beq{\be\begin{array}{c}}
\def\eeq{\end{array}\ee} 
 
\newcommand{\ve}{\varepsilon}

\def\S{{\Omega^{\rm soft}_z}}
\def\dr{{\rm dressed}}

   \makeatletter
  \let\over=\@@over \let\overwithdelims=\@@overwithdelims
  \let\atop=\@@atop \let\atopwithdelims=\@@atopwithdelims
  \let\above=\@@above \let\abovewithdelims=\@@abovewithdelims
\renewcommand\section{\@startsection {section}{1}{\z@}%
                                   {-3.5ex \@plus -1ex \@minus -.2ex}
                                   {2.3ex \@plus.2ex}%
                                   {\normalfont\large\bfseries}}

\renewcommand\subsection{\@startsection{subsection}{2}{\z@}%
                                     {-3.25ex\@plus -1ex \@minus -.2ex}%
                                     {1.5ex \@plus .2ex}%
                                     {\normalfont\bfseries}}

\tikzset{
  on each segment/.style={
    decorate,
    decoration={
      show path construction,
      moveto code={},
      lineto code={
        \path [#1]
        (\tikzinputsegmentfirst) -- (\tikzinputsegmentlast);
      },
      curveto code={
        \path [#1] (\tikzinputsegmentfirst)
        .. controls
        (\tikzinputsegmentsupporta) and (\tikzinputsegmentsupportb)
        ..
        (\tikzinputsegmentlast);
      },
      closepath code={
        \path [#1]
        (\tikzinputsegmentfirst) -- (\tikzinputsegmentlast);
      },
    },
  },
  mid arrow/.style={postaction={decorate,decoration={
        markings,
        mark=at position .5 with {\arrow[#1]{stealth}}
      }}},
}

\tikzset{snake it/.style={decorate, decoration=snake}}
\tikzset{slope/.store in=\slope}
\pgfdeclarelayer{background}
\pgfdeclarelayer{foreground}
\pgfsetlayers{background,foreground}
\newcommand{\theslope}{0.7}
\pgfdeclarepatternformonly{nelines}
{\pgfpoint{-.2mm/\theslope}{-.2mm}}{\pgfpoint{2.2mm/\theslope}{2.2mm}}
{\pgfpoint{2mm/\theslope}{2mm}}
{
    \pgfsetlinewidth{1pt}
    \pgfpathmoveto{\pgfpoint{-.2mm/\theslope}{-.2mm}}
    \pgfpathlineto{\pgfpoint{2.2mm/\theslope}{2.2mm}}
    \pgfusepath{stroke}
}

\begin{document}
\begin{titlepage}
\unitlength = 1mm

\ \\
\vskip 1cm
\begin{center}

{ \LARGE {\textsc{ Infrared Divergences in QED, Revisited}}}

\vspace{0.8cm}
Daniel Kapec$^\dagger$, Malcolm Perry$^*$, Ana-Maria Raclariu$^\dagger$ and Andrew Strominger$^\dagger$

\vspace{1cm}

\begin{abstract}
Recently it has been shown that the vacuum state in QED is infinitely degenerate. Moreover a transition among the degenerate vacua is induced in any nontrivial scattering process and determined from the associated soft factor.  Conventional computations of scattering amplitudes  in QED do not account for this vacuum degeneracy and therefore always give zero. 
This vanishing of all conventional QED amplitudes is usually attributed to infrared divergences. Here we show that if these vacuum transitions are properly accounted for,  the resulting amplitudes are nonzero and infrared finite. 
Our construction of finite amplitudes is mathematically equivalent to, and amounts to a physical reinterpretation of,  the 
1970 construction of Faddeev and Kulish. 
 
  \end{abstract}

\vspace{8.0cm}

\end{center}
\noindent{\it  $^\dagger$ Center for the Fundamental Laws of Nature, Harvard University,
Cambridge, MA, USA}

\noindent{\it $^*$ Department of Applied Mathematics and Theoretical Physics, University of Cambridge, Cambridge,  UK}\\

\end{titlepage}

\pagestyle{empty}
\pagestyle{plain}

\pagenumbering{arabic}

\tableofcontents
\section{Introduction}
Recently it has been shown  \cite{Strominger:2013lka,He:2014cra,Campiglia:2015qka,Kapec:2015ena,He:2015zea} (see \cite{Strominger:2017zoo} for a review) that the infrared (IR) sector of all abelian gauge theories, including QED, is governed by an infinite-dimensional symmetry group.  The symmetry group is generated by large gauge transformations that approach angle-dependent constants at null infinity.  The soft photon theorem is the matrix element of the associated conservation laws. This large gauge symmetry is spontaneously broken, resulting in an infinite 
vacuum degeneracy. 

QED has been tested to 16 decimal places and is the most accurate theory in the history of human thought. The preceding statements have no mathematically new content within QED, and certainly do not imply errors in any previous QED calculations!  However, as emphasized herein,  they do perhaps provide a physically illuminating new way of describing the IR structure.   Moreover,  generalizations of this perspective  to other contexts have led to a variety  of truly new mathematical relations in both gauge theory and gravity \cite{Strominger:2017zoo}.   

One of the puzzling features of the IR structure of QED is the appearance of IR divergences.\footnote{ Originally, these were found by looking at the spectrum and number of photons produced by particles undergoing acceleration. 
Mott \cite{Mott:pcps} looked at corrections to Rutherford scattering as a result of the emission of photons. Bloch and Nordsieck \cite{Bloch:1937pw} examined the spectrum of photons produced by a small change in the velocity of an electron. }
These divergences set all conventional Fock-basis $\cs$-matrix elements to zero. Often they are dealt with by restricting to  inclusive cross sections in which physically unmeasurable photons below some IR cutoff are traced over.  The trace gives a divergence which offsets the zero and yields a finite result for the physical measurement 
\cite{Yennie:1961ad,Kinoshita:1962ur,Lee:1964is,Weinberg:1965nx}. 
While this is adequate for most experimental applications, for many purposes it is nice to have 
an $\cs$-matrix.\footnote{It may also be challenging to describe  experimental measurements of the electromagnetic memory effect 
\cite{Bieri:2013hqa,Susskind:2015hpa,Pasterski:2015zua}  in a theory with a finite IR cutoff. }  For example precise discussions of unitarity or symmetries require an $\cs$-matrix.

It is natural to ask if the newly-discovered IR symmetries are related to the IR divergences of the $\cs$-matrix. We will see that the answer is yes. The conservation laws imply that every non-trivial scattering process is necessarily accompanied by a transition among the degenerate vacua. Conventional QED $\cs$-matrix analyses tend to assume the vacuum is unique and hence that  the initial and final vacua are the same.  Since this violates the conservation laws, the Feynman diagrammatics give a vanishing result. This is usually attributed to `IR divergences', but we feel that this phrase is something of a misnomer. Rather, zero is the correct physical answer. The vanishing of the amplitudes is a penalty for not accounting for  the required vacuum transition. In this paper we allow 
for vacuum transitions to occur, and find that the resulting amplitudes are perfectly IR finite and generically nonvanishing when the conservation laws are obeyed. 

Although we have phrased this result in a way that sounds new, the mathematics behind it is not new. We have merely rediscovered the 1970  formulae \cite{Chung:1965zza,Kibble:1969ip,Kibble:1969ep,Kibble:1969kd,Kulish:1970ut} of Faddeev and Kulish (FK) and others, who showed that certain dressings of charges by clouds of soft photons yield IR finite scattering amplitudes. The FK dressings implicitly generate precisely the required  shift between degenerate vacua. 

While our formulae are not new,\footnote{Except for, in section 6, a conjectured generalization of the FK IR divergence cancellation mechanism  to amplitudes involving some undressed charges but still obeying the conservation laws. An example given there is $e^+e^-$ scattering with no incoming radiation.} our physical interpretation is new. One may hope that the new physical insight will enable a construction of IR finite $\cs$-matrices for unconfined nonabelian gauge theory and also have useful applications to gravity.

Related discussions of the FK construction in the context of large gauge symmetry have appeared in \cite{Gabai:2016kuf,Mirbabayi:2016axw,Gomez:2017soz,Panchenko:2017lkw}.
 
\section{Vacuum selection rules}
In this section we review the derivation of and formulae for the  
vacuum transitions induced by the scattering of charged massless particles. We refer the reader to \cite{Strominger:2013lka,He:2014cra,Mohd:2014oja,He:2015zea} for further details. The conceptually similar  massive case is treated in section \ref{mp}.  Incoming states are best described in advanced coordinates in Minkowski space \begin{equation}
\begin{split}
ds^2 &= -dv^2 +2 dvdr + 2 r^2 \gamma_{z \bz }dz d \bz~,
\end{split}
\end{equation} 
while outgoing states employ retarded coordinates \be ds^2 = -du^2 - 2 dudr + 2 r^2 \gamma_{z \bz }dz d \bz~ .\ee
Here $\gamma_{z \bz }={2\over (1+z\bz)^2}$ is the unit round metric on $S^2$ and $u=t-r$ ($v=t+r$) is the retarded (advanced) time.  The $z$ coordinates used in the advanced and retarded coordinate systems differ by an antipodal map on $S^2$. In and out states are characterized by the charges\footnote{In these and the following equations , ($F_{rv},F_{ru},j_v,j_u$) denote the coefficient of the leading $\co({1 \over r^2})$ term of the large $r$ field expansions, while ($F_{vz},F_{uz}$) denote the leading $\co({ r^0})$ terms. }  \bea Q^-_\ve&=& {1 \over e^2}\int_{\ci^-_+}d^2w \gamma_{w\bw}\ve F_{ru}~,\cr Q^+_\ve&=& {1 \over e^2}\int_{\ci^+_-}d^2w \gamma_{w\bw}\ve F_{rv}~,\eea
where  $F$ is the electromagnetic field strength, $\ci^-_+$ is the future boundary of $\ci^-$, $\ci^+_-$ is the past boundary of $\ci^+$ and 
$\ve$ is any function on $S^2$.  The conservation law for these charges 
\be\label{claw}  \<\o|(Q_\ve^+\cs-\cs Q_\ve^-)|\i\>=0 \ee
is implied by the soft photon theorem. 
In and out soft photon modes are defined as  integrals of the radiative part of
 $F$ over the null generators of past and future null infinity ($\ci^\pm$) according to \be\label{frae}
\int_{-\infty}^\infty du~  F_{u z}\equiv N^+_z~, ~
\ee\be\label{fre}
\int_{-\infty}^\infty dv~  F_{v z}\equiv N^-_z ~.
\ee
Choosing $\ve(w, \bw) ={1 \over z-w}$, \eqref{claw}
can be written in the form 
\be\label{tui} \<\o|(N_z^+\cs-\cs N_z^-)|\i\>=\S\<\o|\cs|\i\>~,\ee
where the soft factor is 
\bea\label{sf} \S=\S^--\S^+ ~, \hspace{70pt}\\ \S^-={e^2 \over 4 \pi}\sum_{k \in \i}{Q_k \over z-z_k}~,\qquad \S^+={e^2 \over 4 \pi}\sum_{k \in \o}{Q_k \over z-z_k} ~.\eea
Here $Q_k$ and $z_k$ denote the charges of the asymptotic particles and the angles at which they enter or exit at $\ci^\pm$. 
Degenerate incoming vacua can be characterized by their $N^-_z$ eigenvalue: 
\be\label{der} N^-_z(z,\bz)| N^\i_z\>= N^\i_z(z,\bz) | N^\i_z\>~.\ee
Let us consider special states denoted $|\i; N^\i_z\>$ comprised of finite numbers of non-interacting incoming charged particles and hard photons built by acting with asymptotic creation operators on 
eigenstates \eqref{der} of $N^-_z$.\footnote{It is assumed here that the Fourier coeffcients of 
the photon creation operators are finite as the frequency $\omega\to 0$.} Such hard particles do not affect the zero modes and hence  obey
\be N^-_z|\i; N^\i_z\>= N^\i_z |\i; N^\i_z\>~.\ee
Adopting a similar notation for out-states, \eqref{tui} becomes
\be\label{tsui} (N^\o_z-N^\i_z)\<\o;N^\o_z|\cs|N_z^\i;\i\>=\S\<\o;N^\o_z|\cs|N_z^\i;\i\>~.\ee
We conclude that either
\be\label{tuei} \<\o;N^\o_z|\cs|N_z^\i;\i\>=0~,\ee
or
\be\label{ui} N^\o_z-N^\i_z=\S ~.\ee
The second relation \eqref{ui} expresses conservation of the charges  - one for each point on the sphere - associated to large gauge symmetries. 
The first states that any amplitude violating the conservation law must vanish. 

In conventional formulations of QED, the vacuum is presumed to be unique.\footnote{Since $N_z$ manifestly carries zero energy, if the vacuum is assumed to be unique it would have to  be an $N_z$ eigenstate.} In that case, \eqref{ui} is not an option, and we conclude that, according to \eqref{tuei}, all  $\cs$-matrix elements vanish. In fact this result is  well-known and attributed  to IR  divergences. We see here that the IR divergences which set all such  amplitudes to zero can be understood as a penalty for neglecting the fact that the in and out vacua differ for every non-trivial scattering process. Armed with this insight, we will construct a natural and IR finite set of scattering amplitudes. 

The necessity for vacuum transitions in any scattering process follows from the constraint equations on 
\ip
\begin{equation}\label{cpeq}
\partial_u F_{ru} + D^z F_{uz} +  D^\bz F_{u\bz}   + e^2   j_u = 0~,
\end{equation} 
and $\ci^-$
\begin{equation}\label{cmeq}
\partial_v F_{rv} - D^z F_{vz} - D^\bz F_{v\bz}   - e^2   j_v = 0~.
\end{equation}
Assuming that the electric field vanishes in the far past and far future and using the matching conditions\footnote{Here we consider theories with no magnetic charges so that $F_{z\bz}|_{\ci^+_-}=0=F_{z\bz}|_{\ci^-_+}$.}
\be\label{mch} F_{ru}|_{\ci^+_-}=F_{rv}|_{\ci^-_+}~,\ee \be\label{mbh}A_{z}|_{\ci^+_-}=A_{z}|_{\ci^-_+}~,\ee
 the divergence of \eqref{ui} (and its complex conjugate) is the sum of the integrals of \eqref{cpeq} and \eqref{cmeq}.  
 
Let us examine the classical electromagnetic field configuration needed to satisfy the constraints.  A single charge $Q_0$ particle  incoming at $(v_0,z_0,\bz_0)$ corresponds to 
 \be j_v=Q_0\delta(v-v_0)\gamma^{z\bz}\delta^{(2)}(z-z_0)~.\ee
 We write the state consisting of one such particle in the $N^\i_z=0$ vacuum as 
  \be |z_0;0\>~,  ~~~~N^-_z|z_0;0\>=0 ~.\ee
 We can solve the constraints for finite $z$
  either using the Coulombic modes with\footnote{$\theta(v)=1$ for $v>0$ and vanishes otherwise. } 

  \be\label{ty}
  F_{rv}= Q_0e^2 \theta(v-v_0)\gamma^{z\bz}\delta^{(2)}(z-z_0)~,\ee
  or with the radiative modes  \be\label{rdr} A_z=-{Q_0 e^2 \over 4\pi}\p_z G(z,z_0)\theta(v-v_0)~,~~~~F_{vz}=-{Q_0 e^2 \over 4\pi}\p_z G(z,z_0)\delta(v-v_0)~,\ee
  where 
  \be \p_z\p_\bz G(z,w)=2\pi\delta^{(2)}(z-w)~.\ee
  For finite $z$ the choice 
  \be\label{we} G(z,w)= \ln|z-w|^2~ \ee
gives  simply 
  \be\label{rdr} A_z=-{Q_0 e^2 \over 4\pi (z-z_0)}\theta(v-v_0)~.\ee The purely Coulombic choice  will violate the matching conditions \eqref{mch} unless the outgoing state also has Coulomb fields at $z=z_0$, where there may not even be any particles on \ip. We first  consider the radiative dressing 
\eqref{rdr}. This potential is pure gauge except at advanced time $v=v_0$ where a radiative shock wave  emerges.  There is a shift in the flat gauge connection between the boundaries 
$\ci^-_+$ and $\ci^-_-$ of $\mathcal{I}^-$ given by $N^-_{z}=-{Q_0 e^2 \over 4\pi (z-z_0)}.$

Of course more general solutions of the constraints, which do involve Coulomb fields, are possible and can be obtained by adding to \eqref{rdr} any solution of the source-free equation. Indeed the difference between \eqref{ty} and \eqref{rdr} is such a solution. We will return to the more general case in sections 5 and 6. 

The Green function $G$ in \eqref{we} leads to image charges at $z=\infty$. Since there are no physical charges presumed at this point  and we must preserve the constraints, delta function `wires' of non-zero $F_{vr}$ are added  connecting the images at various values of $v_k$ where particles enter. One such wire with net integral $\sum_{k\in \i}Q_k$ will cross to \ip.  Overall charge conservation guarantees, if a similar construction is used to satisfy the \ip\ constraints,  that this will match with the  $z=\infty$ wire on \ip. 

Of course these wires can be smoothed out by adding source-free solutions of the free Maxwell equation.\footnote{There may be a preferred Lorentz covariant dressing if the charged particles are taken as conformal primaries rather than plane waves as in \cite{Pasterski:2016qvg}.} 
For example  we can use \be\label{dep} G(z,w)=\ln\big[ |z-w|^2(1+z\bz)^{-1}(1+w\bw)^{-1}\bigr]~,\ee
which obeys $2\p_z\p_\bz G=4\pi \delta^{(2)}(z-w) - \gamma_{z\bz}.$
This effectively spreads the image charges, and along with them the $F_{vr}$ flux wires required for their cancellation, evenly over the sphere.

For our purposes we are primarily interested in the structure near 
$z=z_0$ which has the same singularity for all the $G$'s. The choice of $G$ will not be central and we focus on the simplest one \eqref{we}.

\section{Dressed quantum states}

The story of dressed charges began with Dirac \cite{Dirac:1955uv} who realized that part of the problem with the formulation of quantum electrodynamics was that conventional states for charged particles were not gauge invariant. Suppose one is considering the Dirac field for an electron, $\psi(x)$. Then one usually thinks of the operator $\psi(x)$ as creating an electron at the point $x$. In classical physics, if one makes a gauge transformation
\begin{equation} \label{gtg} A_\mu \rightarrow A_\mu + \partial_\mu \varepsilon(x)~,
\end{equation}
then the gauge transformation of a field with charge $Q_0$ is 
\begin{equation} \label{gte} \psi(x) \rightarrow e^{iQ_0\varepsilon(x)}\psi(x)~.
\end{equation}
Dirac made a field invariant under gauge transformations which die at infinity by introducing a dressing of the charged particle. One
replaces $\psi(x)$ by the gauge invariant $\psi^*(x)$ defined by
\begin{equation} \label{gaugeinv} \psi^*(x) = \psi(x)e^{iQ_0\int A^\mu(x^\prime)\, C_\mu(x^\prime)\, d^4x^\prime}~.
\end{equation}
$C_\mu(x)$ is then required to obey the equation
\begin{equation} \label{ceqn} \partial_\mu C^\mu = \delta^{(4)}(x-x^\prime)~.\end{equation}
Solutions to this equation are of course not unique, since we can add to it any solution of the homogeneous equation. Thus Dirac's prescription is not unique, and may involve either radiative or Coulomb modes depending on how $C^\mu$ is chosen. 

In quantum field theory in the Schr\"odinger picture, operators are time independent and so one would
replace these expressions by the corresponding non-covariant forms in which only the spatial components of $A_\mu $ and $C^\mu$ are used. The integral is then taken over a three-dimensional spatial section
of spacetime and the four-dimensional delta-function is replaced by the three-dimensional delta function.
Thus 
\begin{equation} \label{gaugeinvs} \psi^*(x) = \psi(x)e^{iQ_0\int A^i(x^\prime)\, C_i(x^\prime)\, d^3x^\prime} \end{equation}
and
\begin{equation} \label{ceqns} \partial_i C^i = \delta^{(3)}(x-x^\prime)~.\end{equation}

This prescription replaces the bare electron by an electron together with an electromagnetic cloud. 
It is important to note that the operator (\ref{gaugeinv}) is only invariant under small gauge transformations vanishing sufficiently quickly at infinity, since an integration by parts is needed in order to demonstrate invariance. Under large gauge transformations, Dirac's operators transform with a phase, as charged operators should.
Dirac provided an illuminating example of a $C_i$ that satisfies $\partial_iC^i = \delta^{(3)}(x-x^\prime)$.
This is just
\begin{equation} C_i = -\partial_i \Biggl\{\frac{1}{4\pi\vert x-x^\prime\vert}\Biggr\}~.\end{equation}
$C_i$ is then just the electric Coulomb field of a point charge. 

In the full interacting quantum theory, the radiative modes of the electromagnetic field obey the exact $\ci^-$ commutator
\be\label{ccr} \bigl[A_w(v,w,\bw),F_{v'\bz}(v',z,\bz)\bigr]={ie^2\over 2}\delta(v-v')\delta^{(2)}(w-z)~. \ee
The commutators of Coulombic modes are then, according to Dirac, whatever they must be  in order that the constraints \eqref{cmeq} are satisfied. That is 
the operator $F_{rv}$ is defined by 
\begin{equation}
\begin{split}\label{cmq}
 F_{rv}\equiv\int_{-\infty}^vdv'\big(D^z F_{v'z} + D^\bz F_{v'\bz}  + e^2   j_{v'}\big)~,
\end{split}
\end{equation}
where the constant of integration is set (for massless charges only) by demanding that the Coulomb field vanish in the far past. 
Its commutators are then computed using \eqref{ccr} along with those for the matter fields appearing in $j_v$.

The coherent quantum state corresponding to \eqref{rdr} is, up to a large gauge transformation, 
\bea\label{kli} |z_0;0\>_\dr &\equiv& e^{iR_0} |z_0;0\>~,\cr 
R_0&\equiv&\frac{Q_0}{2\pi}\int d^2w\gamma_{w \bw} G(z_0,w)D\cdot A(v_0,w,\bw)~,\eea
where $D\cdot A\equiv D^wA_w+D^\bw A_\bw$. 
We may describe this as a charged particle surrounded by a cloud of soft photons. It easily follows from \eqref{ccr} and \eqref{cmq} that $ |z_0;0\>_\dr$ obeys the constraints without any Coulombic $F_{vr}$ wires extending out of the charge, and the matching condition \eqref{mch} is trivially satisfied.   
The dressing shifts the action of $F_{vz}$ on states by 
\be [F_{vz}(v,z,\bz),iR_0] =-{Q_0 e^2\delta(v-v_0)\over 4\pi(z-z_0)}~.\ee
We see explicitly that the early and late vacua on $\ci^-$ differ by a large gauge transformation and
\be  N_z^-|z_0; 0\>_\dr= -{Q_0 e^2\over 4\pi(z-z_0)}|z_0; 0\>_\dr~.\ee
Had we started with a vacuum state with nonzero $N^\i_z$ the dressing would have simply shifted the eigenvalue. 

The dressed single particle state \eqref{kli} is easily generalized to a multiparticle state
\bea\label{kl} |\i; 0\>_\dr &\equiv& e^{iR} |\i; 0\>~,\cr R&\equiv& \frac{1}{2\pi}\int dv d^2w\gamma_{w \bw} d^2z \gamma_{z\bz}j_v(v,z,\bz)G(z,w)D\cdot A(v,w,\bw)~ .\eea
Also for outgoing states 
\be\label{kl}\< \o;  0|_\dr \equiv \< \o;  0|e^{-iR}~.\ee
The dressed states accompany the charges with nonzero eigenvalues for the soft photon operator, 
\be 
\begin{split} N^-_z |\i; 0\>_\dr &=-\sum_{k \in \i}{Q_k e^2 \over 4\pi(z-z_k)}|\i; 0\>_\dr~,  \\
\< \o;  0|_\dr N^+_z &= - \< \o;  0|_\dr\sum_{k \in \o}{Q_k e^2 \over 4\pi(z-z_k)}~.\end{split} \ee
In particular, the eigenvalues automatically obey the selection rule \eqref{ui}
\be \label{support}N^+_z-N^-_z=\S~, \ee
so that generically
\be\label{iy}  \< \o;  0|_\dr \cs |\i; 0\>_\dr \neq 0~. \ee

\tikzstyle{edges}=[very thick]
\tikzstyle{particles}=[very thick, blue]
\tikzstyle{interaction}=[circle, draw=black, pattern=nelines, very thick, inner sep=0pt, minimum size = 1.5cm]
\tikzstyle{positrons}=[red, very thick]
\colorlet{photons}{yellow!90!red!8!white}
\colorlet{phcont}{yellow!90!red}
\tikzstyle{int}=[circle, draw=black, pattern=nelines, very thick, inner sep=0pt, minimum size = 1.5cm]
\tikzstyle{cloudl}=[cloud, draw=phcont,cloud puffs=8,cloud puff arc=120, aspect=2, inner ysep=.5em, very thick, fill=photons, rotate=25, minimum size=.8cm]
\tikzstyle{cloudr}=[cloud, draw=phcont,cloud puffs=8,cloud puff arc=120, aspect=2, inner ysep=.5em, very thick, fill=photons, rotate=-25, minimum size=0.8cm]
\tikzstyle{hph}=[very thick, snake=snake, color=yellow]
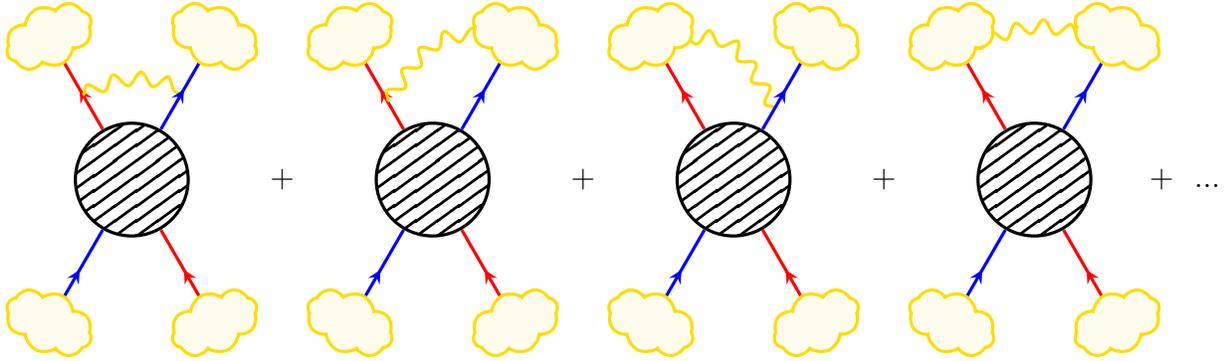
\begin{figure}
\begin{tikzpicture}
[node distance=2cm]
\begin{pgfonlayer}{foreground}
\node[int] (int1) at (0,0) {};
\node[draw=white] (p1) [right of=int1] {$+$};
\node[int] (int2) [right of=p1] {};
\node[draw=white] (p2) [right of=int2] {$+$};
\node[int] (int3) [right of=p2] {};
\node[draw=white] (p3) [right of=int3] {$+$};
\node[int] (int4) [right of=p3] {};
\node[draw=white] (p4) [right of=int4] {$+ ~~...$};
\node [cloudl] (c1ul) at (120:2.2cm) {};
\node [cloudr] (c1ur) at (60:2.2cm) {};
\node [cloudl] (c1dr) at (-60:2.2cm) {};
\node [cloudr] (c1dl) at (-120:2.2cm) {};
\node [cloudl] (c2ul) at ($(int2) +(120:2.2)$) {};
\node [cloudr] (c2ur) at ($(int2) +(60:2.2)$) {};
\node [cloudl] (c2dr) at ($(int2) +(-60:2.2)$) {};
\node [cloudr] (c2dl) at ($(int2) +(-120:2.2)$) {};
\node [cloudl] (c2ul) at ($(int3) +(120:2.2)$) {};
\node [cloudr] (c2ur) at ($(int3) +(60:2.2)$) {};
\node [cloudl] (c2dr) at ($(int3) +(-60:2.2)$) {};
\node [cloudr] (c2dl) at ($(int3) +(-120:2.2)$) {};
\node [cloudl] (c2ul) at ($(int4) +(120:2.2)$) {};
\node [cloudr] (c2ur) at ($(int4) +(60:2.2)$) {};
\node [cloudl] (c2dr) at ($(int4) +(-60:2.2)$) {};
\node [cloudr] (c2dl) at ($(int4) +(-120:2.2)$) {};
\end{pgfonlayer}
\begin{pgfonlayer}{background}
\path [draw, positrons, postaction={on each segment={mid arrow=red}}] (int1) -- (120:2cm);
\path [draw, positrons, postaction={on each segment={mid arrow=red}}] (-60:2cm) -- (int1);
\path[draw=phcont, snake it, very thick] (120:1.3cm) arc (120:60:1.3cm);
\path [draw, particles, postaction={on each segment={mid arrow=blue}}] (int1) -- (60:2cm);
\path [draw, particles, postaction={on each segment={mid arrow=blue}}] (-120:2cm) -- (int1);
\path [draw, positrons, postaction={on each segment={mid arrow=red}}] (int2) -- +(120:2cm);
\path [draw, positrons, postaction={on each segment={mid arrow=red}}] ($(int2) +(-60:2cm)$) -- (int2);
\path[draw=phcont, snake it, very thick] ($(int2) +(120:1.2cm)$) ..controls ($(int2) +(90:1.9cm)$).. ($(int2) +(60:2.1cm)$);
\path [draw, particles, postaction={on each segment={mid arrow=blue}}] (int2) -- +(60:2cm);
\path [draw, particles, postaction={on each segment={mid arrow=blue}}] ($(int2) +(-120:2cm)$) -- (int2);
\path [draw=blue, positrons, postaction={on each segment={mid arrow=red}}] (int3) -- +(120:2cm);
\path [draw=blue, positrons, postaction={on each segment={mid arrow=red}}] ($(int3) +(-60:2cm)$) -- (int3);
\path[draw=phcont, snake it, very thick] ($(int3) +(60:1.1cm)$) ..controls ($(int3) +(90:1.9cm)$).. ($(int3) +(120:2cm)$);
\path [draw, particles, postaction={on each segment={mid arrow=blue}}] (int3) -- +(60:2cm);
\path [draw, particles, postaction={on each segment={mid arrow=blue}}] ($(int3) +(-120:2cm)$) -- (int3);
\path [draw=blue, positrons, postaction={on each segment={mid arrow=red}}] (int4) -- +(120:2cm);
\path [draw=blue, positrons, postaction={on each segment={mid arrow=red}}] ($(int4) +(-60:2cm)$) -- (int4);
\path[draw=phcont, snake it, very thick] ($(int4) +(120:2cm)$) ..controls ($(int4) + (90:2.1cm)$)..($(int4) +(60:2cm)$);
\path [draw, particles, postaction={on each segment={mid arrow=blue}}] (int4) -- +(60:2cm);
\path [draw, particles, postaction={on each segment={mid arrow=blue}}] ($(int4) +(-120:2cm)$) -- (int4);
\end{pgfonlayer}
\end{tikzpicture}
\caption{IR divergences arise from soft photon exchange between pairs of external charges. When the charges are dressed with apprpropriately correlated clouds of soft photons, these divergences are pairwise cancelled by exchanges involving the soft clouds. }
\label{fig1}
\end{figure}  
In fact the dressed amplitudes are free of IR divergences altogether. The basic mechanism is illustrated in figure \ref{fig1}.  IR divergences arise from the exchange of soft photons between pairs of external legs. These exponentiate in such a way to cause  ordinary Fock-basis amplitudes to vanish. However, when the charged particles are dressed by soft photon clouds, further divergences arise when a soft photon is exchanged between one external leg and the soft cloud surrounding the second external leg or between the pair of soft clouds. 
FK \cite{Kulish:1970ut} showed that by a judicious choice of such a soft cloud one can arrange for the IR divergences to cancel and obtain an IR finite $\cs$-matrix. In the next section we show that our dressed states differ from those of FK only by terms which are subleading in the IR, and therefore effect the same IR cancellations. 

\section{FK states}

Faddeev and Kulish  \cite{Kulish:1970ut}, building on Dirac and others \cite{Chung:1965zza,Kibble:1969ip,Kibble:1969ep,Kibble:1969kd}, developed a scheme for dressing charged particles which eliminates IR divergences. Their starting point was to argue that the LSZ procedure for identifying asymptotic states is inapplicable in quantum electrodynamics: since the electromagnetic interaction has infinite range, there can be no isolated interaction region. They resolved this by observing that the action for charged particles contains a term
\begin{equation} \label{current}
\int J^\mu\, A_\mu\, d^4x
\end{equation}
where $J^\mu$ is the electromagnetic current. Since this current is conserved, the action is gauge invariant provided appropriate boundary conditions hold for the gauge transformations. This is a special case of Dirac's treatment which leads to a collection of soft photons accompanying any charged particle. 

If one studies the state for a single electron of three-momentum $p^i$, then in the eikonal approximation the current is the classical current of a single charged particle located at
$x^i={p^it/m}$.
FK  dressed a single particle charged state $|\vec p \>$ with the associated soft cloud 
\be\label{dew} |\vec p \>_{FK}=\exp\left[-{eQ_0\over (2\pi)^3}\int {d^3q\over 2 q_0}\big(f^\mu a^\dagger_\mu(\vec q)-f^{*\mu} a_\mu(\vec q)\big)\right]|\vec p \>~,~~~~~~f^\mu=\left[ {p^\mu\over p\cdot q}-c^\mu \right]e^{i{p\cdot k \over p_0}t}~,\ee
where $c^\mu$ satisfies $c\cdot q=1, \; c^2=0$. In fact, they demonstrate that in order to cancel infrared divergences it is sufficient to choose an arbitrary dressing
\be 
f^\mu=\left[ {p^\mu\over p\cdot q}-c^\mu \right]\psi(p,q)~,
\ee
with the condition that $\psi(p,q)=1$ in a neighborhood of $q=0$. 
For a multi-particle state with zero net total charge and minimal dressing $\psi=1$, the $c^\mu$ terms cancel out of the dressing function and we can deal solely with the dressing factor
\be\label{fdr}
f^\mu= {p^\mu\over p\cdot q}~.
\ee

We wish to rewrite \eqref{dew} in terms of asymptotic fields at $\ci$. 
The plane-wave expansion of the radiative modes of the electromagnetic potential is given by
\begin{equation}
\begin{split}\label{modexp}
A_\nu (x) = e \sum_{\alpha=\pm} \int \frac{d^3q}{(2\pi)^3} \frac{1}{2\omega} \left[ \ve_\nu^{*\alpha} (\vec{q}) a_\alpha^{\text{in}} (\vec{q}) e^{i q \cdot x }  
+ \ve_\nu^{\alpha} (\vec{q}) a_\alpha^{\text{in}} (\vec{q})^\dagger e^{- i q \cdot x }  \right]~,
\end{split}
\end{equation}
where $q^2=0$, the two polarization vectors satisfy a normalization condition $\ve^\nu_\alpha \ve^*_\beta{}_\nu  = \delta_{\alpha \beta }$ and 
\begin{equation}
\begin{split}
\left[ a_\alpha^{\text{in}} (\vec{q}\hspace{2pt}) , a_\beta^{\text{in}} (\vec{q}\hspace{2pt}')^\dagger \right] = \delta_{\alpha\beta} (2\pi)^3   2 \omega \delta^{(3)} \left( \vec{q} - \vec{q}\hspace{2pt}' \right) ~ .
\end{split}
\end{equation}
Null momenta can be characterized by a point on the asymptotic $S^2$ and an energy $\omega$
\begin{equation}\label{nullmomentum}
\begin{split}
q^\mu = \frac{\omega}{1+z\bz} \left( 1+ z\bz, z+\bz,-i(z-\bz),1-z\bz\right) = (\omega, q^1, q^2, q^3)~.
\end{split}
\end{equation}
Now use the expansion for the spatial part of the plane-wave wavefunctions,
\begin{equation} e^{i\vec{q}\cdot \vec{x}} = \sum_l i^l\,(2l+1)\,j_l(qr)\,P_l(\cos\gamma) \end{equation}
where $j_l$ are the spherical Bessel functions, $\gamma$ is the angle between $\vec{q}$ and $\vec{x}$ and $q=| \hspace{.02in} \vec{q}\hspace{.02in} |$.
The asymptotic form of $j_l(qr)$ for $qr \gg 1$ is 
\begin{equation} j_l(qr) \sim \frac{1}{qr} \sin(qr-{\frac{1}{2}}\pi l) \end{equation}
which yields an approximation that localizes the momenta of the gauge field in the optical direction \cite{Strominger:2017zoo}
\begin{equation}\label{sx1}
\begin{split}
A_z^{(0)} (v,z,\bz) &= \lim_{r \to \infty} A_z(v,r,z,\bz) \\
&= - \frac{i}{8\pi^2}  \frac{\sqrt{2} e }{1+ z \bz }  \int_0^\infty d\omega
 \left[  a_+^{\text{in}} ( \omega {\hat x} ) e^{-i \omega v }   - a_-^{\text{in}} ( \omega {\hat x} )^\dagger e^{i \omega v }   \right]  ~,~~~~~\hat x  = \hat x (z, \bz)~.
\end{split}
\end{equation}
One then finds 
\be  |\vec p \>_{FK}=
\exp\left[ {i\over 2\pi} \int dv d^2w\gamma_{w \bw}d^2z \gamma_{z\bz}j_v(v,z,\bz) G(z,w)D\cdot A(0,w,\bw)\right] |\vec p\>~.\ee
This is almost the same as our dressed state \eqref{kli}, except that the soft cloud in an FK state is always at $v=0$, whereas in \eqref{kli} it appears at the same advanced time $v_0$ as the charged particle.\footnote{ The FK states can still satisfy the constraints, at the price of Coulomb fields extending from $v=0$ to the locations of the charged particles. } This difference is subleading in 
$\omega$, since leading order quantities at $\omega=0$ are insensitive to null separations. In fact, with the natural choice
\be
\psi=e^{i\omega v_0}~,
\ee
which clearly satisfies $\psi=1$ near $\omega=0$, one obtains precisely the dressing in section 3. Since the two dressings differ only by terms higher order in $\omega$,
both implement the all-orders cancellation of IR divergences established by FK. 

\section{Massive Particles}
\label{mp}
In this section the discussion is generalized to massive particles. 
The soft factor $\S$, which determines the change in the vacuum state, is given for massless particles  in \eqref{sf} in terms of the points $z_k$ at which they exit or enter the celestial sphere. Such a formula cannot exist for massive particles in eigenstates with   momentum $p^\mu_k$  as they never reach null infinity.  Instead a massive particle is characterized by a point on the unit 3D hyperboloid $H_3$ which may be parameterized by \be \hat p^\mu={p^\mu \over m}~,~~~\hat p^2=-1~.\ee
As described in \cite{Campiglia:2015qka,Campiglia:2015lxa}, the contribution to $\S$ from such a particle is proportional to
\be\label{ds} G_z(\hat p_k)=\int d^2w G(w,\bw; \hat p_k){1 \over w-z}~,\ee
where $G$ here is the bulk-to-boundary propagator on $H_3$ 
obeying $\square G=0$.  If we infinitely boost $\hat p$, $G$ reduces to a boundary delta function.  Hence, in analogy with \eqref{kli},  in order to prevent IR divergences from setting the amplitudes to zero, we should dress such massive particle states 
as\bea\label{kldi} |\hat p_1,\ldots;0\>_\dr &\equiv& e^{iR_m} |\hat p_1,\ldots;0\>~, \cr R_m&\equiv& \sum_{k\in \i}\frac{Q_k}{2\pi}\int d^2z(G_z (\hat p_k)A_\bz(0,z,\bz)+h.c.)~,\eea 
where to avoid separate discussion of the zero mode we restrict to 
the special case $\sum_{k\in \i}Q_k=0$ with zero net charge.
It is straightforward to show that this is precisely the  FK state 
given by \eqref{fdr} and therefore has IR finite scattering amplitudes. 

\begin{figure}
\centering
\begin{tikzpicture}[scale = 5]
\begin{pgfonlayer}{foreground}
\draw[style=edges] (0,0) node[left] {} -- (1,1) node[right] {\hspace{5pt}$\mathcal{I}^+_+$} -- node[midway, above] {\hspace{40pt} {\large $\mathcal{I}^+$}} (2,0) node[above] {\hspace{10pt}$\mathcal{I}^+_-$} node[below]{\hspace{10pt}$\mathcal{I}^-_+$} -- node[midway, below] {\hspace{40pt}{\large $\mathcal{I}^-$}} (1,-1) node[right] {\hspace{5pt}$\mathcal{I}^-_-$} -- cycle;
\node[interaction] (int) at (1,0) {};
\path [draw, particles, postaction={on each segment={mid arrow=blue}}] (int) .. controls +(-.25,.6)  ..(1,1);
\path [draw, positrons, postaction={on each segment={mid arrow=red}}] (int) .. controls +(.15,.6) ..(1,1);
\path [draw=blue, particles, postaction={on each segment={mid arrow=blue}}] (int) .. controls +(-.1,.6) ..(1,1);
\path [draw, positrons, postaction={on each segment={mid arrow=red}}] (1,-1) .. controls +(.25,.4) ..(int);
\path [draw, positrons, postaction={on each segment={mid arrow=red}}] (int) .. controls +(-.15,.6) .. (1,1);
\path [draw=blue, particles, postaction={on each segment={mid arrow=blue}}] (1,-1) .. controls +(-.25,.4) ..(int);
\draw[-> ,snake=snake, line after snake=1mm, very thick, color=phcont] (int) -- +(-.35,.35); 
\draw[snake=snake, line before snake=1mm, very thick, color=phcont] (.65,.35) -- (.5,.5); 
\draw[snake=snake, line after snake=.5mm, very thick, color=phcont] (int) -- +(-.35,-.35);
\draw[->, snake=snake, line after snake=1.5mm, very thick, color=phcont]  (.5,-.5) -- (.65,-.35); 
\draw[->, snake=snake,line after snake=1mm, very thick, color=phcont] (int) -- +(.35,.35);
\draw[snake=snake, line before snake=1mm, very thick, color=phcont] (1.35,.35) -- (1.5,.5); 
\draw[snake=snake, line after snake=.5mm, very thick, color=phcont] (int) -- +(.35,-.35);
\draw[->, snake=snake, line after snake=1.5mm, very thick, color=phcont] (1.5,-.5) -- (1.35,-.35); 
\draw[|-|] (.45,.55) -- node[rotate=45, fill=white]{$F_{ru}\neq 0$} (.95,1.05);
\draw[|-|] (.45,-.55) -- node[rotate=-45, fill=white]{$F_{rv}\neq 0$} (.95,-1.05);
\draw[|-|] (-.05,.05) -- node[rotate=45, fill=white]{$F_{ru}= 0$} (.45,.55);
\draw[|-|] (-.05,-.05) -- node[rotate=-45, fill=white]{$F_{rv} = 0$} (.45,-.55);
\end{pgfonlayer}
\end{tikzpicture}
\caption{Dressed massive particles of zero net charge come in from $i^-$, scatter and go out to $i^+$. The Faddeev-Kulish dressing introduces radiative shock waves at $v=0$ and $u=0$ which cancel the asymptotic Coulomb fields of the particles for $v>0$ and $u<0$ respectively.
For neutral scattering states the Coulomb field will vanish near spatial infinity while for charged ones it will be an angle-dependent constant. }
\label{fig2}
\end{figure}
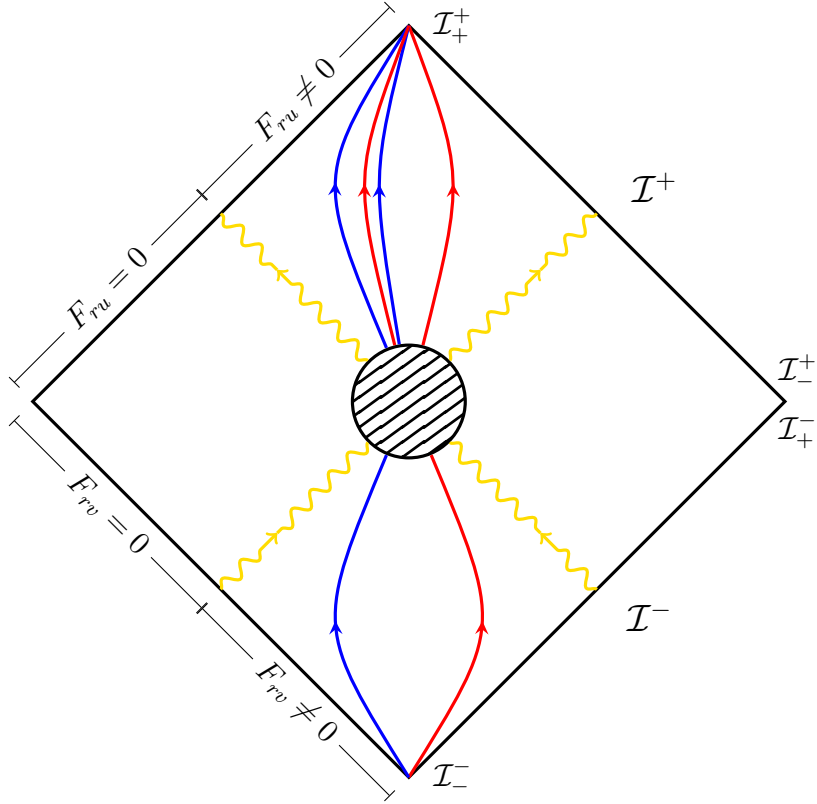

Unlike the massless case studied above, this construction gives  Coulomb fields. One finds
\be\label{kdi} [D^zF_{vz}+D^\bz F_{v\bz},iR_m] = -\delta(v)\gamma^{z\bz}\sum_{k\in \i}Q_k e^2 G(z,\bz,\hat p_k) ~.\ee 
This is a radiative shock wave coming out at $v=0$.  As there are no charged particles incoming at $v=0$, 
the constraints then imply that $F_{rv}(z,\bz)$ must shift by $-\gamma^{z\bz}\sum_{k\in \i}Q_k e^2 G(z,\bz,\hat p_k)$ at $v=0$. This is precisely the (negative of the) asymptotic incoming Coulomb field in the absence of any radiation,  associated with a collection of incoming massive point particles with momentum $m\hat p_k$. The constant part of $F_{rv}$ is fixed by demanding that near $\ci^-_-$ it equal the Coulomb field sourced by the massive charges entering through past timelike infinity $i^-$. This gives 
\be F_{rv}=\theta(-v)\gamma^{z\bz}\sum_{k\in \i}Q_k e^2 G(z,\bz,\hat p_k)~.\ee
The  effect of the radiative shock wave is to set to zero the Coulomb fields after $v=0$ (note we are considering zero net charge). Similarly, fixing the integration function for $F_{ru}$ with a boundary condition at $\ci^+_+$,  the corresponding out state has no Coulomb fields before $u=0$. This is illustrated in figure \ref{fig2}.
Since we then have 
\be\label{retd} F_{ru}|_{\ci^+_-}=0=F_{rv}|_{\ci^-_+}~,\ee
 this construction guarantees that the matching condition \eqref{mch}
 is trivially satisfied and  the amplitude need not vanish. Had we not restricted to the zero charge sector,  the boundary field strengths \eqref{retd} would be angle-independent constants. 
 
 Note that for massive charged particles in plane wave states, the soft photon cloud can never be `on top of' the particle. Massive particles  go to timelike infinity, while radiative photons always disperse to null infinity.  The simplest FK states have them coming out at $u=0$, but exactly when they come out, or the fact that they come out before the charges themselves is unimportant since only the leading IR behavior of the cloud is relevant to the cancellation of divergences. 

\section{Charged states} 
In this section we consider a more general class of physical states in which Coulomb fields persist
to $\ci^+_-$ and $\ci^-_+$, the generic charges are nonzero and the matching conditions nontrivially satisfied. 

The condition \eqref{retd}, which is obeyed by all FK states (with zero net charge)  implies that the in and out charges $Q^\pm_\ve$ vanish.  Explicitly one finds  
\be Q^-|_{\ve={1 \over z-w}}|\text{in};0\rangle_{\text{dressed}} =\left({4\pi \over e^2}N^-_z - \sum_{k\in in} Q_k G_z(\hat{p}_k)\right)|\text{in}; 0\rangle_{\text{dressed}} =0~. \ee
\be \langle \text{out}; 0|_{\text{dressed}} Q^+|_{\ve={1 \over z-w}}=\langle \text{out}; 0|_{\text{dressed}} \left({4\pi \over e^2}N^+_z -\sum_{k\in out} Q_k G_z(\hat{p}_k) \right)=0~.\ee
That is the incoming (outgoing) soft photon cloud shields all of the nonzero mode charges of the incoming (outgoing) 
charged particles. Since the charges all commute with $a^{\i \dagger}_\pm(\vec k)$, adding radiative photons will not change the charges.\footnote{Assuming  the frequency distribution does not have poles or other singularities for $\omega \to 0$. If it does have such poles, new IR singularities may appear, and the state would not be among those shown by FK to be IR finite. }

In contrast to the situation for FK states, nonzero mode $Q^\pm_\ve$ charges are generically all nonvanishing in the real world. Consider for example  $e^+e^-$ scattering in the center-of-mass frame with incoming velocities $\pm \vec v$, unaccompanied by incoming radiation. 
The coefficient of the ${1\over r^2}$ radial electric field is given by the Li\'enard-Wiechert formula
\be\label{LWf} F_{rv}={e^2(1-\vec v^2)  \over 4\pi \big(1+\hat x \cdot \vec v \big)^2}-{e^2(1-\vec v^2)  \over 4\pi \big(1-\hat x \cdot \vec v \big)^2}=-\frac{e^2}{\pi}\frac{\hat{x}\cdot \vec{v}(1- \vec{v}^2)}{(1-(\hat{x}\cdot \vec{v})^2)^2}\ .\ee
The charges constructed from this are nonzero :  \be Q^-|_{\ve={1 \over z-w}}= -\int_{\ci^{-}_{+}} d^2w \gamma_{w\bar{w}}\frac{1}{z-w}  \frac{1}{\pi} \frac{\hat{x}\cdot \vec{v}(1- \vec{v}^2)}{(1-(\hat{x}\cdot \vec{v})^2)^2} = Q^+|_{\ve={1 \over z-w}}=G_z(\hat{p}_-)-G_z(\hat{p}_+) \ , \ee
where $\hat{p}_-$ $(\hat{p}_+)$ is the momentum of the electron (positron). 
Non-zero $Q^\pm_\ve$ charges can be generically sourced by radiative Maxwell fields and arise even in the absence of charged particles. Source free initial data for the Maxwell equation is given by specifying an arbitrary function 
$F_{vz}(v,z,\bz)$ on $\ci^-$. Assuming that $F_{rv}$ vanishes in the far past, one has 
\be \label{zro} F_{rv}(z,\bz)|_{\ci^-_+} =\int_{-\infty}^\infty dv \left(D^z F_{vz}(v,z,\bz) + D^{\bz}F_{v\bz}(v,z,\bz)\right)\ .\ee
Demanding that the right hand side vanishes is a nonlocal constraint on the incoming initial data. 
Such nonlocal constraints are indeed imposed on FK states, which are dressed charged particles plus radiative modes. 
As mentioned above the frequency-space coefficients of the field operators $A_z$  are (except for the charge dressings)  presumed to be finite for  $\omega \to 0$, which precisely imposes the nonlocal constraint on the field strength that the integral \eqref{zro} vanish. 

Quantum states which describe these physical situations with nonzero $Q^\pm_\ve$ charges certainly exist, even if they are not FK states.\footnote{We note that all the non-zero mode charges can be shielded, classically or quantum mechanically,  by a correlated cloud of very soft radiation with arbitrarily small energy at arbitrarily large radius. In this sense any state can be approximated by an FK state.}   It is natural to ask if such states can ever have IR finite scattering amplitudes. Given our earlier argument that the true role of IR divergences is simply to enforce conservation of all the charges, one might expect it to be possible.  We now propose  that this is indeed the case.

The basic idea is that the soft photon clouds and charged particles can be separated without affecting the IR cancellation mechanism of FK, even if we  move a particle from incoming to outgoing, leaving its cloud intact. However moving a charged particle from incoming to outgoing will in general take a zero-charge FK state to one with all charges excited. 

Let's consider Bhabha scattering as a specific example 
\be e^+e^-\to e^+e^- ~,\ee
where the incoming and out going charges are all given FK dressings.
Then there will be an incoming wave of photons shielding the incoming charges at $v=0$ and an outgoing one at $u=0$ shielding the outgoing charges. The long range fields will be angle-independent, in contrast to \eqref{LWf}.
The scattering is IR finite, with IR divergences from soft photon exchanges between pairs of external charges canceled by divergences from soft photon exchanges between external particles  and radiative clouds and pairs of radiative clouds. This was depicted in figure 1. 

\begin{figure}
\begin{tikzpicture}
[node distance=2cm]
\begin{pgfonlayer}{foreground}
\node[int] (int1) at (0,0) {};
\node[draw=white] (p1) [right of=int1] {$+$};
\node[int] (int2) [right of=p1] {};
\node[draw=white] (p2) [right of=int2] {$+$};
\node[int] (int3) [right of=p2] {};
\node[draw=white] (p3) [right of=int3] {$+$};
\node[int] (int4) [right of=p3] {};
\node[draw=white] (p4) [right of=int4] {$+~~...$};
\node [cloudl] (c1ul) at (120:2.2cm) {};
\node [cloudr] (c1ur) at (60:2.2cm) {};
\node [cloudl] (c1dr) at (-60:2.2cm) {};
\node [cloudr] (c1dl) at (-120:2.2cm) {};
\node [cloudl] (c2ul) at ($(int2) +(120:2.2)$) {};
\node [cloudr] (c2ur) at ($(int2) +(60:2.2)$) {};
\node [cloudl] (c2dr) at ($(int2) +(-60:2.2)$) {};
\node [cloudr] (c2dl) at ($(int2) +(-120:2.2)$) {};
\node [cloudl] (c2ul) at ($(int3) +(120:2.2)$) {};
\node [cloudr] (c2ur) at ($(int3) +(60:2.2)$) {};
\node [cloudl] (c2dr) at ($(int3) +(-60:2.2)$) {};
\node [cloudr] (c2dl) at ($(int3) +(-120:2.2)$) {};
\node [cloudl] (c2ul) at ($(int4) +(120:2.2)$) {};
\node [cloudr] (c2ur) at ($(int4) +(60:2.2)$) {};
\node [cloudl] (c2dr) at ($(int4) +(-60:2.2)$) {};
\node [cloudr] (c2dl) at ($(int4) +(-120:2.2)$) {};
\end{pgfonlayer}
\begin{pgfonlayer}{background}
\path [draw, particles, postaction={on each segment={mid arrow=blue}}] (210:2.5cm) -- (int1);
\path [draw=blue, positrons, postaction={on each segment={mid arrow=red}}] (-60:2cm) -- (int1);
\path[draw=phcont, snake it, very thick] (210:1.3cm) arc (210:60:1.3cm);
\path [draw, particles, postaction={on each segment={mid arrow=blue}}] (int1) -- (60:2cm);
\path [draw, particles, postaction={on each segment={mid arrow=blue}}] (-120:2cm) -- (int1);
\path [draw, hph] (int1)--(35:2cm);
\path [draw, particles, postaction={on each segment={mid arrow=blue}}] ($(int2) +(210:2.5cm)$) -- (int2);
\path [draw=blue, positrons, postaction={on each segment={mid arrow=red}}] ($(int2) +(-60:2cm)$) -- (int2);
\path[draw=phcont, snake it, very thick] ($(int2) +(210:1.2cm)$) ..controls ($(int2) +(180:1.5cm)$) and ($(int2) +(110:1.9cm)$).. ($(int2) +(60:2.1cm)$);
\path [draw, particles, postaction={on each segment={mid arrow=blue}}] (int2) -- +(60:2cm);
\path [draw, particles, postaction={on each segment={mid arrow=blue}}] ($(int2) +(-120:2cm)$) -- (int2);
\path [draw, hph] (int2)-- +(35:2cm);
\path [draw, particles, postaction={on each segment={mid arrow=blue}}] ($(int3) +(210:2.5cm)$) -- (int3);
\path [draw=blue, positrons, postaction={on each segment={mid arrow=red}}] ($(int3) +(-60:2cm)$) -- (int3);
\path[draw=phcont, snake it, very thick] ($(int3) +(60:1.1cm)$) ..controls ($(int3) +(90:1.9cm)$).. ($(int3) +(120:2cm)$);
\path [draw, particles, postaction={on each segment={mid arrow=blue}}] (int3) -- +(60:2cm);
\path [draw, particles, postaction={on each segment={mid arrow=blue}}] ($(int3) +(-120:2cm)$) -- (int3);
\path [draw, hph] (int3)-- +(35:2cm);
\path [draw, particles, postaction={on each segment={mid arrow=blue}}] ($(int4) +(210:2.5cm)$) -- (int4);
\path [draw=blue, positrons, postaction={on each segment={mid arrow=red}}] ($(int4) +(-60:2cm)$) -- (int4);
\path[draw=phcont, snake it, very thick] ($(int4) +(120:2cm)$) ..controls ($(int4) + (90:2.1cm)$)..($(int4) +(60:2cm)$);
\path [draw, particles, postaction={on each segment={mid arrow=blue}}] (int4) -- +(60:2cm);
\path [draw, particles, postaction={on each segment={mid arrow=blue}}] ($(int4) +(-120:2cm)$) -- (int4);
\path [draw, hph] (int4)-- +(35:2cm);
\end{pgfonlayer}
\end{tikzpicture}
\caption{In this figure, the outgoing positron in Figure 1 is crossed to an incoming electron, but its associated soft photon cloud remains as part of the out state.  The leading  IR divergences from the depicted soft photon exchange still cancel, even though the in and out states carry nontrivial $Q^\pm_\ve$ charges and are no longer Faddeev-Kulish states.} \label{fig3}
\end{figure}
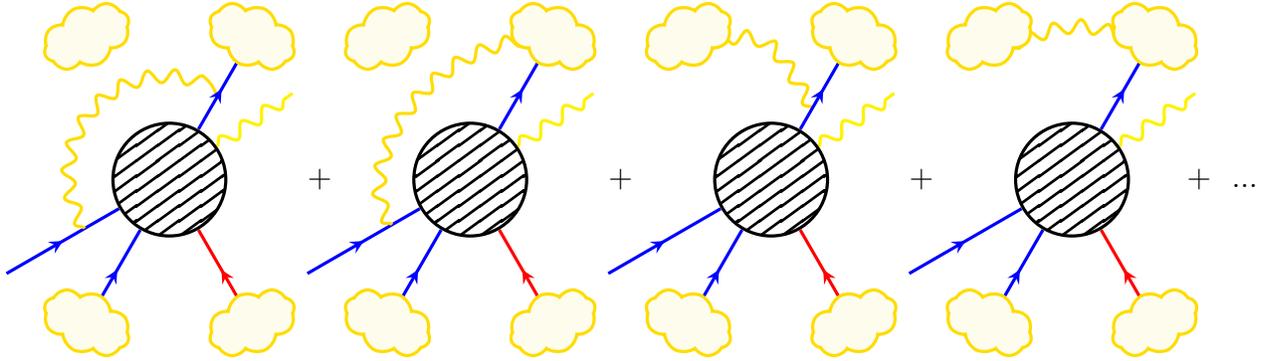

Now let us  move the outgoing positron to an ingoing electron with the  same momentum, and add a radiative photon to the out state to conserve energy and momentum: 
\be e^+e^-e^-\to e^-+\gamma ~ .\ee This has no effect on the soft factor, since the motion from out to in and  the change in the sign of the charge each contribute a factor minus one.  The same $(-1)^2=1$ applies  to the leading IR divergence of an attached soft photon and  ensures that these soft exchanges will continue to cancel. See figure \ref{fig3}. 

Since we have also done nothing to $N_z^\pm$, the conservation law   (\ref{support})  remains satisfied. However we have changed the charges. While they were previously zero, the contribution from the FK shield of the outgoing positron is no longer cancelled, while the new incoming electron does not have an FK  shield.  These give equal contributions to the incoming and outgoing charges 
\be Q^\pm(z)= G_z(\hat{p})~. \ee
Ultimately one might hope to use crossing symmetry to  prove IR finiteness in this context. FK IR cancellations occur order by order in Feynman diagram  perturbation theory, while crossing symmetry 
also holds in perturbation theory. The action of crossing a single out particle to an in one produces general $Q^\pm_\ve$  charges while 
changing  FK to some more general class of states. We conjecture that  scattering amplitudes among  these more general charged states are IR finite.

\section*{Acknowledgements}
The authors are  grateful  to David Gross, Sasha Haco, Zohar Komargodski, Kumar Narain, Massimo Porrati, Burk Schwab and Sasha Zhiboedov  for useful conversations. This work is supported by DOE grant DE-SC0007870 and the United Kingdom STFC.

\end{document}